\documentclass[a4paper,11pt]{article}
\usepackage{pos}

\title{The $B_c$ lifetime in the Standard Model}
\author*[a]{Jason~Aebischer}
\author[b]{Benjam\'in Grinstein}

\affiliation[a]{Physik-Institut, Universit\"at Z\"urich, CH-8057 Z\"urich, Switzerland}

\affiliation[b]{Department of Physics, University of California at San Diego,
    La Jolla, CA 92093, USA}

\emailAdd{jason.aebischer@physik.uzh.ch}
\emailAdd{bgrinstein@ucsd.edu}

\abstract{Using an operator product expansion (OPE) approach an updated Standard Model prediction of the $B_c$ lifetime is presented. The computation in three different mass schemes for the heavy quarks leads to three different values consistent with each other and with experiment. Furthermore a novel way to compute the $B_c$ lifetime is presented, taking differences of $B,D$ and $B_c$ meson decay rates. In this approach the leading contributions from free-quark decays cancel out, leading to a reduction of scale and scheme dependence.}

\FullConference{%
  *** Particles and Nuclei International Conference - PANIC2021 ***\\
  *** 5 - 10 September, 2021 ***\\
  *** Online ***
}

\begin{document}
\maketitle

\section{Introduction}

The lifetime of the $B_c=(\overline b c)$ meson is an interesting observable, as it allows to put strong constraints on New Physics models such as for instance scalar Leptoquarks or Two-Higgs-Doublet models \cite{Alonso:2016oyd,Blanke:2018yud,Blanke:2019qrx}. Such models are particularly interesting regarding the charged-current $B$-anomalies seen in $R_{D}$ and $R_{D^*}$. The lifetime of the $B_c$ is measured with a small uncertainty by the LHCb \cite{LHCb:2014ilr,LHCb:2014glo} and CMS \cite{CMS:2017ygm} collaborations and averages to
\begin{equation}\label{eq:exp}
  \tau_{B_c}^{\text{exp}} = 0.510(9)\text{ps}\,.
\end{equation}
From the theory side, different approaches can be employed to determine the $B_c$ lifetime, such as QCD Sum Rules \cite{Kiselev:2000pp}, Potential models \cite{Gershtein:1994jw} or Operator Product Expansion (OPE) methods \cite{Beneke:1996xe,Bigi:1995fs,Chang:2000ac}. The three approaches all predict $\tau_{B_c}$ to be in the ballpark of the experimental value in Eq.~\eqref{eq:exp}. In the following we will adopt the approach introduced by Beneke and Buchalla in \cite{Beneke:1996xe} to compute $\tau_{B_c}$ using Effective Field Theory methods involving the OPE, the effective Hamiltonian as well as Non-Relativistic QCD (NRQCD). It can be summarized as follows: In a first step the heavy degrees of the Standard Model (SM) are integrated out at the electroweak scale to obtain the effective Hamiltonian $\mathcal{H}_{\text{eff}}$ contributing to the $B_c$ lifetime. In a next step, the Wilson coefficients of the effective operators in $\mathcal{H}_{\text{eff}}$ are run down to the low scale $\mu_\text{low}$ by solving their renormalization group equations. This low scale is chosen to be of the order of the $b$-quark mass for $\overline b$-decays and of order $m_c$ for $c$-quarks decays. At these scales an OPE of the transition operator $\mathcal{T}_{B_c}$ relevant for the $B_c$ decay is performed. The transition operator is related via the optical theorem to the decay rate of the $B_c$ in the following way:

\begin{equation}
  \Gamma_{B_c} = \frac{1}{2M_{B_c}}\langle B_c|\mathcal{T}_{B_c}|B_c\rangle\,,
\end{equation}
where the transition operator is the imaginary part of the time-ordered product of two insertions of the effective Hamiltonian

\begin{equation}
  \mathcal{T}_{B_c} = \text{Im}\, i\,\int d^4x\, T\, \mathcal{H}_{\text{eff}}(x)\mathcal{H}_{\text{eff}}(0)\,.
\end{equation}

The products of two operators from $\mathcal{H}_{\text{eff}}$ at different space-time points are then expanded in the OPE, i.e. written in terms of a series of local operators.
The relevant contributions for $\mathcal{T}_{B_c}$ can be split into contributions resulting from the $\overline b$- and $c$-quark decays, as well as from Weak Annihilation (WA) and Pauli interference (PI) diagrams:

\begin{equation}
  \mathcal{T}_{B_c} = \mathcal{T}_{\overline b}+\mathcal{T}_{c}+\mathcal{T}_{\text{WA}}+\mathcal{T}_{\text{PI}}\,.
\end{equation}
The first two contributions result from spectator decays of the corresponding quarks inside the $B_c$ meson, generating the free-quark decay contribution (at dimension three) as well as dimension-five chromomagnetic dipole operators. The WA and PI contributions correspond to dimension-six operators in the OPE. In spite of their higher mass suppression in the expansion their contributions are retained, since they are generated through one-loop diagrams and hence have an enhancement factor of 16$\pi^2$ compensating the suppression.

After performing the OPE at the low scale $\mu_\text{low}$ the QCD fields are expressed in terms of NRQCD fields. Subsequently, a velocity expansion in the small quark velocities of the non-relativistic quarks is performed. The matrix elements of the resulting operators can be estimated for the leading operators like the kinetic term, the Fermi- and the Darwin term using potential models \cite{Gershtein:1994dxw}. In estimating the matrix elements of the four-quark operators spin-symmetry has been used, which relates all matrix elements of the four-fermi operators from WA and PI to a single reduced matrix element.

\section{Result}
In this section we present the results for the final decay rates of the $B_c$, which are discussed in detail in \cite{Aebischer:2021ilm}. Assuming a non-zero strange quark mass and computing in three different mass schemes, namely the MSbar scheme, the meson scheme and the Uspilon scheme we find for the decay rates

\begin{equation}\label{eq:schemes}
\begin{aligned}
  \Gamma^{\overline{\text{MS}}}_{B_c} &= (1.51\pm 0.38|^{\mu}\pm 0.08|^{\text{n.p.}}\pm 0.02|^{\overline{m}}  \pm0.01|^{m_s}\pm 0.01|^{V_{cb}})\,\,\text{ps}^{-1}\,, \\
  \Gamma^{\text{meson}}_{B_c} &= (1.70\pm 0.24|^{\mu}\pm 0.20|^{\text{n.p.}} \pm0.01|^{m_s}\pm 0.01|^{V_{cb}})\,\,\text{ps}^{-1} \,,  \\
  \Gamma^{\text{Upsilon}}_{B_c} &= (2.40\pm 0.19|^{\mu}\pm 0.21|^{\text{n.p.}} \pm0.01|^{m_s}\pm 0.01|^{V_{cb}})\,\,\text{ps}^{-1} \,.
\end{aligned}
\end{equation}

The largest uncertainties in all three schemes stem from residual dependence on the renormalization scale $\mu$, which can be reduced by taking into account higher-order QCD corrections. Further uncertainties result from non-perturbative (n.p.) corrections. These can be reduced by taking into account higher order corrections in the velocity expansion as well as by having better estimates for the matrix elements, preferably from Lattice calculations. The smallest uncertainties are due to parametric uncertainties of the strange-quark mass and the CKM element $V_{cb}$. When the strange quark mass is neglected, the central values of the decay rates are enhanced by $\sim 7\%$.

When combining all the different uncertainties the results in the three schemes in Eq.~\eqref{eq:schemes} are compatible with each other and also with the experimental value, which is derived from Eq.~\eqref{eq:exp} to be

\begin{equation}\label{eq:Gexp}
  \Gamma_{B_c}^\text{exp} = 1.961(35) \,\text{ps}^{-1}\,.
\end{equation}

The large spread in the central values obtained using different mass schemes calls however for a computation which includes higher-order QCD- as well as non-perturbative corrections in order to decrease the apparent differences.

\section{New method to determine $\Gamma_{B_c}$}
In this section we describe a novel way \cite{Aebischer:2021eio} on how to determine the $B_c$ decay rate, using differences of heavy meson decay rates. Generally, the decay rate of a heavy meson $H$ with heavy quark $Q$ can be written as
\begin{equation}\label{eq:GM}
  \Gamma(H_Q) = \Gamma_Q^{(0)}+\Gamma^{n.p.}(H_Q)+\Gamma^{\text{WA}+\text{PI}}(H_Q)+\mathcal{O}(\frac{1}{m_Q^4})\,,
\end{equation}
where the leading contribution $\Gamma_Q^{(0)}$ denotes the free quark decay rate, $\Gamma^{n.p.}$ results from non-perturbative corrections and $\Gamma^{\text{WA,PI}}$ denote the WA and PI contributions. Taking now the difference of decay rates for the $B$, $D$ and $B_c$ mesons, and applying the formula in Eq.~\eqref{eq:GM} one finds
\begin{align}\label{eq:diff}
  \Gamma(B)+\Gamma(D)-\Gamma(B_c) &= \Gamma^{n.p.}(B)+\Gamma^{n.p.}(D)-\Gamma^{n.p.}(B_c) \nonumber \\
  &+\,\Gamma^{\text{WA}+\text{PI}}(B)+\Gamma^{\text{WA}+\text{PI}}(D)-\Gamma^{\text{WA}+\text{PI}}(B_c)\,.
\end{align}

The right-hand side of Eq.~\eqref{eq:diff} can be computed in a similar way as discussed in the introduction with the only difference that for the $B$ and $D$ mesons Heavy Quark Effective Theory instead of NRQCD is employed. On the left-hand side of Eq.~\eqref{eq:diff} the decay rates $\Gamma(B)$ and $\Gamma(D)$ can be taken from experiment, which then allows to express $\Gamma(B_c)$ in terms of known/calculable quantities.

The main advantage of Eq.~\eqref{eq:diff} is that the free-quark decay contribution drops out in the difference together with it's uncertainties. Furthermore, one can use this relation for either charged or neutral mesons, leading to four possible ways to predict the $B_c$ decay rate. For the meson scheme and the four different combinations of mesons used in the relation we report the results in Tab.~\ref{tab:res}.
The obtained results again show some deviation compared to the experimental value in Eq.~\eqref{eq:Gexp}. Several possibilities might serve as a solution to resolve this discrepancy: A first explanation would be the underestimation of the uncertainties from NLO corrections to the Wilson coefficients as well as non-perturbative corrections in our computation. Secondly eye-graph contributions, which are generally neglected in lattice computations of the used matrix elements, might have a sizable impact on the result. Furthermore, as pointed out in \cite{King:2021xqp} neglected dimension-seven contributions in the charm decays can have a large impact. Finally quark-hadron duality might be violated, in which case a completely new method to compute the decay rates of heavy mesons might be in order.

\begin{table}[t]
\centering
 \begin{tabular}{|l |c |c |c |c|}
 \hline
 & $B^0,D^0$ & $B^+,D^0$ & $B^0,D^+$ & $B^+,D^+$ \\ [0.5ex]
 \hline \hline
$\Gamma^{\text{meson}}_{B_c}$ & 3.03 $\pm$ 0.51 & 3.03 $\pm$ 0.53 & 3.33 $\pm$ 1.29 & 3.33 $\pm$ 1.32 \\
 \hline
 \end{tabular}
 \caption{\small
Results obtained for the decay rate $\Gamma(B_c)$ in $\text{ps}^{-1}$ in the meson scheme, using the different combinations of $B$ and $D$ mesons in Eq.~\eqref{eq:diff}.
}
  \label{tab:res}
\end{table}

\section{Summary}
We present an updated computation of the $B_c$ lifetime in the SM using an EFT approach involving the effective Hamiltonian, the OPE as well as NRQCD. The obtained values in three different mass schemes are compatible with each other and with the experimental value within the given uncertainties. The large scheme dependence of the result, manifesting itself in a large spread of the central values ranging from 1.51 $\text{ps}^{-1}$ to 2.40 $\text{ps}^{-1}$, calls however for a determination including higher order QCD corrections.

Furthermore a novel way on how to determine the $B_c$ lifetime is presented, taking differences of $B,\,D$ and $B_c$ decay rates. The method leads to values that exceed the experimental value, which might have several reasons, like the underestimation of uncertainties, the neglected eye-graph and dimension-seven terms or even the violation of quark-hadron duality.

\small
\bibliographystyle{unsrt}
\bibliography{AGbib}

\end{document}